\documentclass[aps,prb,preprint,superscriptaddress,showpacs,showkeys,floatfix]{revtex4}

\usepackage{graphicx,color}

\newcommand{\jwj}[1]{\textcolor{red}{#1}}

\graphicspath{{figs/}}
\begin{document}
\title{Mechanical Properties of MoS$_{2}$/Graphene Heterostructures}
\author{Jin-Wu Jiang}
    \altaffiliation{Corresponding author: jwjiang5918@hotmail.com}
    \affiliation{Shanghai Institute of Applied Mathematics and Mechanics, Shanghai Key Laboratory of Mechanics in Energy Engineering, Shanghai University, Shanghai 200072, People's Republic of China}
\author{Harold S. Park}
    \altaffiliation{Corresponding author: parkhs@bu.edu}
    \affiliation{Department of Mechanical Engineering, Boston University, Boston, Massachusetts 02215, USA}

\date{\today}
\begin{abstract}
We perform classic molecular dynamics simulations to comparatively investigate the mechanical properties of single-layer MoS$_{2}$ and a graphene/MoS$_{2}$/graphene heterostructure under uniaxial tension. We show that the lattice mismatch between MoS$_{2}$ and graphene will lead to an spontaneous strain energy in the interface. \jwj{The Young's modulus of the heterostructure is much larger than that of MoS$_{2}$. While the stiffness is enhanced, the yield strain of the heterostructure is considerably smaller than the MoS$_{2}$ due to lateral buckling of the outer graphene layers owning to the applied mechanical tension.}
\end{abstract}

\pacs{68.65.Ac, 62.25.-g}
\keywords{Molybdenum Disulphide, Heterostructure, Mechanical Property}
\maketitle
\pagebreak

\section{introduction}

Molybdenum disulphide (MoS$_{2}$) is a semiconductor with a bulk band gap of about 1.2~{eV},\cite{KamKK} which can be further manipulated by reducing its thickness to monolayer, two-dimensional form.\cite{MakKF} This finite band gap is a key reason for the excitement surrounding MoS$_{2}$ as compared to another two-dimensional material, graphene, as graphene is well-known to be gapless.\cite{NovoselovKS2005nat}  Because of its direct band gap and also its properties as a lubricant, MoS$_{2}$ has attracted considerable attention in recent years.\cite{WangQH2012nn,ChhowallaM,RadisavljevicB2011nn,HuangW,JiangJW2013mos2,VarshneyV,JiangJW2013sw,JiangJW2013bend,BertolazziS,CooperRC2013prb1,CooperRC2013prb2,gomezAM2012,Castellanos-GomezA2013nl} Although graphene intrinsically has zero band gap, it is the strongest material in the nature with Young's modulus above 1.0~TPa.\cite{LeeC2008sci} Furthermore, graphene possesses a superior thermal conductivity that may be useful in removing heat from electronic devices.\cite{BalandinAA2008}

The key point from the above discussion is that MoS$_{2}$ and graphene have complementary physical properties.  Therefore, it is natural to investigate the possibility of combining graphene and MoS$_{2}$ in specific ways to create heterostructures that mitigate the negative properties of each individual constituent.\cite{BritnellL2013sci,RoyaK2013ssc,Algara-SillerG2013apl,ZanR2013acsn,MyoungN2013acsn,BertolazziS2013nl,LarentisS2014nl,ZhangW2014sr} For example, graphene/MoS$_{2}$/graphene (GMG) heterostructures have better photon absorption and electron-hole creation properties, because of the enhanced light-matter interactions by the single-layer MoS$_{2}$.\cite{BritnellL2013sci} Another experiment recently showed that MoS$_{2}$ can be protected from radiation damage by coating it with graphene layers, which is a design that exploits the outstanding mechanical properties of graphene.\cite{ZanR2013acsn} Although experimentalists have shown great interest in the mechanical properties of GMG heterostructures, the corresponding theoretical efforts have been quite limited until now. 

Therefore, the objective of the present work is to present an initial theoretical investigation of the mechanical properties of GMG heterostructures.  We perform classical molecular dynamics (MD) simulations to comparatively study the mechanical properties of single-layer MoS$_{2}$ and the GMG heterostructure. We first point out a spontaneous strain energy arising at the MoS$_{2}$/graphene interface, which results from the mismatch between the lattice constants of MoS$_{2}$ and graphene. We find that the Young's modulus of MoS$_{2}$ can be greatly increased by sandwiching it between two outer graphene layers. However, our simulations also illustrate that the yield strain in the MoS$_{2}$ is reduced significantly due to buckling of the outer graphene layers.

\section{structure and simulation details}

All MD simulations in this work were performed using the publicly available simulation code LAMMPS~\cite{PlimptonSJ}, while the OVITO package was used for visualization~\cite{ovito}. The standard Newton equations of motion were integrated in time using the velocity Verlet algorithm with a time step of 1~{fs}. Periodic boundary conditions were applied in the two in-plane directions, while free boundary conditions were applied in the out-of-plane direction. The structure is uniaxially stretched with a strain rate of $\dot{\epsilon}=10^{9}$~s$^{-1}$, which is a typical value in MD simulations as shown in previous works.~\cite{JiangJW2012jmps} \jwj{The strain is applied in the armchair direction of graphene and MoS$_{2}$. The structure is deformed in x-direction, while keeping the other direction stress free.}

\begin{figure}[htpb]
  \begin{center}
    \scalebox{1.0}[1.0]{\includegraphics[width=8cm]{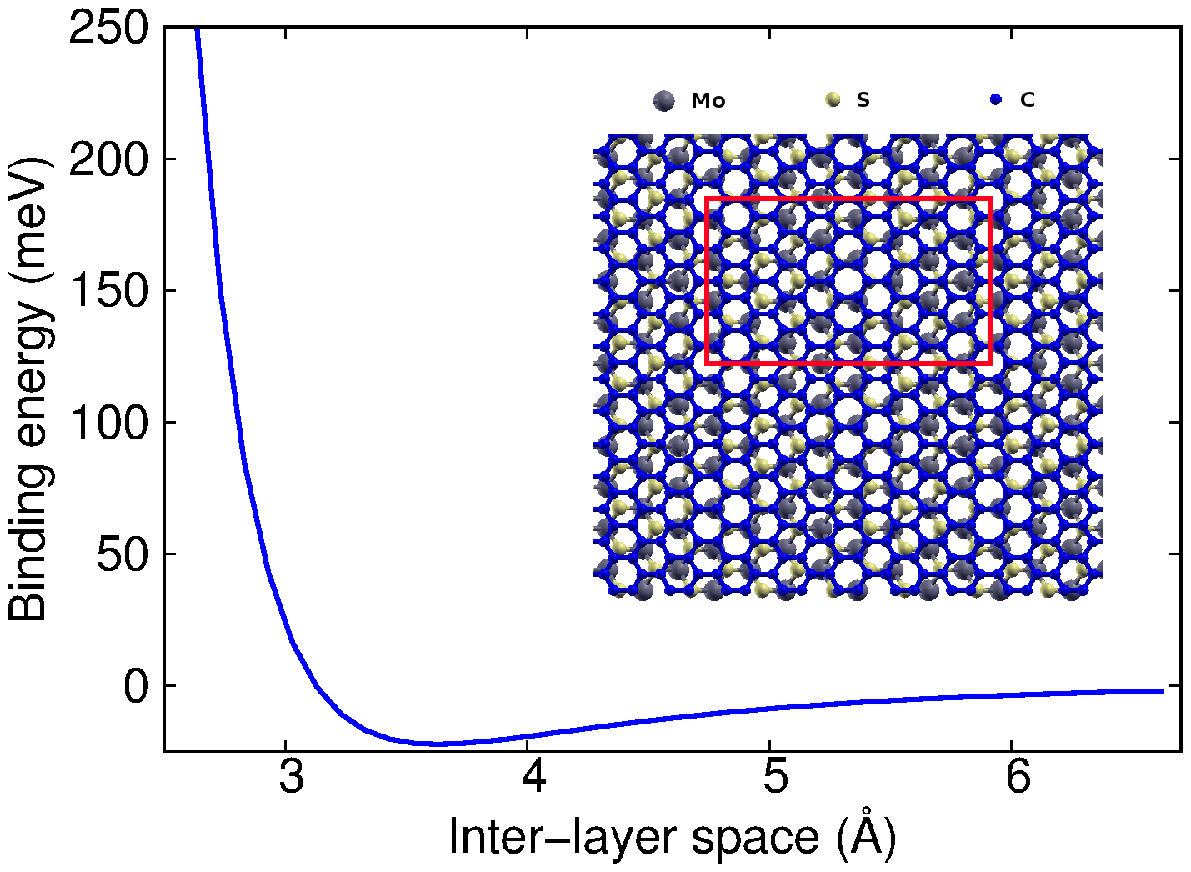}}
  \end{center}
  \caption{(Color online) Binding energy (per carbon atom) between the single-layer MoS$_{2}$ and graphene. Inset shows the top view of the MoS$_{2}$/graphene heterostructure. The red box ($21.61\times 12.48$~{\AA}) displays a translation supercell for the heterostructure.}
  \label{fig_binding_energy_per_carbon}
\end{figure}

The MoS$_{2}$ interatomic interactions are described by a recently developed Stillinger-Weber potential,\cite{JiangJW2013sw} while the carbon-carbon interactions are described by the second generation Brenner (REBO-II) potential~\cite{brennerJPCM2002}. The MoS$_{2}$ and graphene layers in the GMG heterostructure are coupled by van der Waals interactions, which are described by the Lennard-Jones potential. The energy and distance parameters in the Lennard-Jones potential are $\epsilon$=3.95~{meV} and $\sigma$=3.625~{\AA}, while the cutoff is 10.0~{\AA}. These potential parameters are determined by fitting to the interlayer spacing and the binding energy between a single-layer of MoS$_{2}$ and a single-layer of graphene. Fig.~\ref{fig_binding_energy_per_carbon} shows the energy vs. interlayer spacing between MoS$_{2}$ and graphene. The potential energy minimum is found at the interlayer spacing of 3.63~{\AA}, with a corresponding binding energy of -22.357~{meV}. These two values are very close to the first-principle predictions around -21.0~{meV} and 3.66~{\AA} in Ref.~\onlinecite{MiwaRH2013jpcm}, or -23.0~meV and 3.32~{\AA} in Ref.~\onlinecite{MaY2011ns}. Fig.~\ref{fig_binding_energy_per_carbon} inset shows the top view of the GMG heterostructure.

\section{results and discussion}
\jwj{The lattice constants are 2.49~{\AA} and 3.12~{\AA} for graphene and MoS$_{2}$, respectively. The size of the unit cell in the armchair direction is 4.31~{\AA} and 5.40~{\AA} for graphene and MoS$_{2}$, respectively. It is true that, due to their different lattice constants, there will always be some lattice mismatch and strain energy in both graphene and MoS$_{2}$. However, the size of the graphene with 5 unit cells in the armchair direction is about 21.55~{\AA}, which is almost the same as the size of the MoS$_{2}$ with 4 unit cells in the armchair direction, i.e 21.60~{\AA}. That is graphene and MoS$_{2}$ can be perfectly matched when 5 graphene unit cells are on top of 4 MoS$_{2}$ unit cells. After optimization, we find that the size of this supercell is about 21.61~{\AA}, which is shown in the inset of Fig.~\ref{fig_binding_energy_per_carbon}. Similarly, the size of the supercell in the zigzag direction is about 12.48~{\AA}}. Hence, the dimension of the supercell is $21.61\times 12.48$~{\AA}.  Therefore, an ideal super lattice (without spontaneous strain) can be constructed if the heterostructure is obtained by duplicating the supercell in the two-dimensional plane. However, some mechanical strain will be introduced in the heterostruce, if its dimension is not exactly an integer times the supercell. In this situation, one atomic layer is stretched while its neighboring layer will be compressed, so that these two neighboring layers become the same size. We refer to this strain energy as the spontaneous intrinsic strain energy in the heterostructure. 

\begin{figure}[htpb]
  \begin{center}
    \scalebox{1.0}[1.0]{\includegraphics[width=8cm]{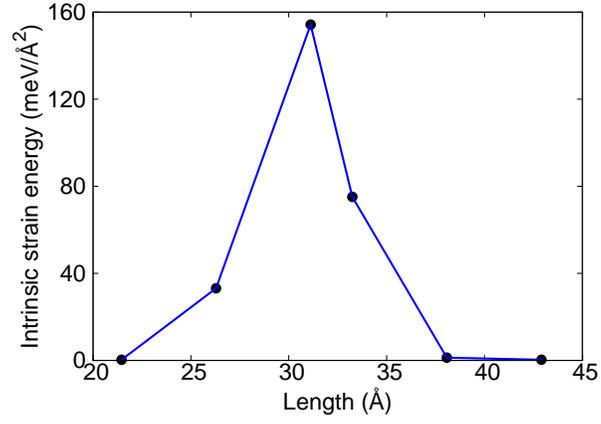}}
  \end{center}
  \caption{(Color online) Intrinsic strain energy density in the MoS$_{2}$/graphene heterostructure of different length. The width of the heterostructure is 12.48~{\AA}.}
  \label{fig_intrinsic_strain_energy}
\end{figure}

Fig.~\ref{fig_intrinsic_strain_energy} shows the length dependence for the spontaneous intrinsic strain energy in the GM heterostructure with a fixed width of 12.48~{\AA}, which is exactly the width of the supercell. The spontaneous intrinsic strain energy density ($E_{\rm is}$) in the figure is calculated by following equation,$E_{\rm is} = ( E_{\rm GM} - E_{G} - E_{M} ) / A$, where $A$ is the area of the GM interface and $E_{\rm GM}$, $E_{G}$, and $E_{M}$ are the potential energies in the GM heterostructure, single-layer graphene, and the single-layer MoS$_{2}$, respectively. The van der Waals interaction between the graphene and the MoS$_{2}$ is not included in this calculation, so that the resulted value is purely the strain energy stored in the graphene or MoS$_{2}$ layer. The spontaneous intrinsic strain energy is zero at the two boundaries with $L_{x}=21.61$ and 43.22~{\AA}. These value are one or two times the supercell size shown in the inset of Fig.~\ref{fig_binding_energy_per_carbon}. It confirms that there is indeed no strain energy for the ideal GM heterostructure. The spontaneous intrinsic strain energy reaches a maximum value in the heterostructure with length around one and half times the supercell size.

Having established the supercell dimensions, we now continue to perform a comparative study on the mechanical behavior of single-layer MoS$_{2}$ and the GMG heterostructure of dimension $64.82\times 49.90$~{\AA} under uniaxial tension. There are $3\times 4$ supercells in the GMG heterostructure, so there is no spontaneous intrinsic strain energy in this examined structure. During the tensile loading of the heterostructure, both graphene and MoS$_{2}$ layers are stretched simultaneously.

\begin{figure}[htpb]
  \begin{center}
    \scalebox{1.0}[1.0]{\includegraphics[width=8cm]{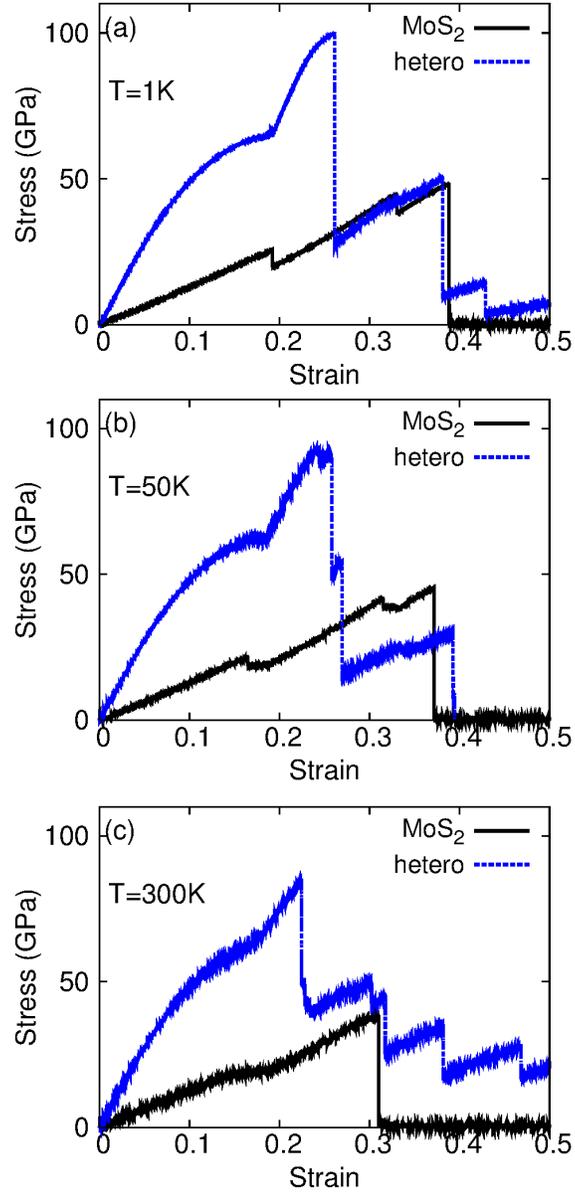}}
  \end{center}
  \caption{(Color online) Stress-strain relation in the MoS$_{2}$ and the GMG heterostructure. The temperature is 1.0~K in (a), 50.0~K in (b), and 300.0~K in (c).}
  \label{fig_stress_strain}
\end{figure}

We first compare the Young's modulus in the single-layer MoS$_{2}$ and the GMG heterostructure. Fig.~\ref{fig_stress_strain} shows the stress-strain relationship in these two systems at three temperatures 1.0~K, 50.0~K, and 300.0~K.  \jwj{Thickness is not a well-defined quantity in one-atomic thick layered materials such as graphene and MoS$_{2}$. Hence, we have assumed the thickness of the single-layer graphene to be the space between two neighboring graphene layers in the three-dimensional graphite. Similar technique is applied for the thickness of the single-layer MoS$_{2}$. That is the thickness is 3.35~{\AA} and 6.09~{\AA} for single-layer graphene and MoS$_{2}$, respectively.} \jwj{The mechanical strength of the GMG heterostructure is considerably larger than the MoS$_{2}$.} The strength enhancement is due to the high strength of single-layer graphene, whose Young's modulus is around 1.0~TPa.\cite{LeeC2008sci} The Young's modulus ($Y$) of the GMG heterostructure can be predicted by the following rule of mixtures based on the arithmetic average,\cite{KarkkainenKK}
\begin{eqnarray}
Y_{\rm GMG} = Y_{G}f_{G} + Y_{M}f_{M},
\end{eqnarray}
where $Y_{\rm GMG}$, $Y_{G}$, and $Y_{M}$ are the Young's modulus for GMG heterostructure, graphene, and MoS$_{2}$, respectively. $f_{G}=2V_G / (2V_G + V_M)=0.524$ is the volume fraction for the two outer graphene layers in the GMG heterostructure, and \jwj{$f_{M}=V_M / (2V_G + V_M)=0.476$} is the volume fraction for the inner MoS$_{2}$ layer. In our simulations, the room temperature Young's modulus are 859.69~GPa for graphene and 128.75~GPa for MoS$_{2}$. From this mixing rule, the upper-bound Young's modulus of the GMG heterostructure is 511.76~GPa. The Young's modulus for the GMG heterostructure at room temperature is 556.33~GPa, which is higher than the value predicted by the mixing rule because of the interlayer van der Waals interaction between graphene and MoS$_{2}$ layers in the GMG heterostructure.

\begin{figure}[htpb]
  \begin{center}
    \scalebox{1.0}[1.0]{\includegraphics[width=8cm]{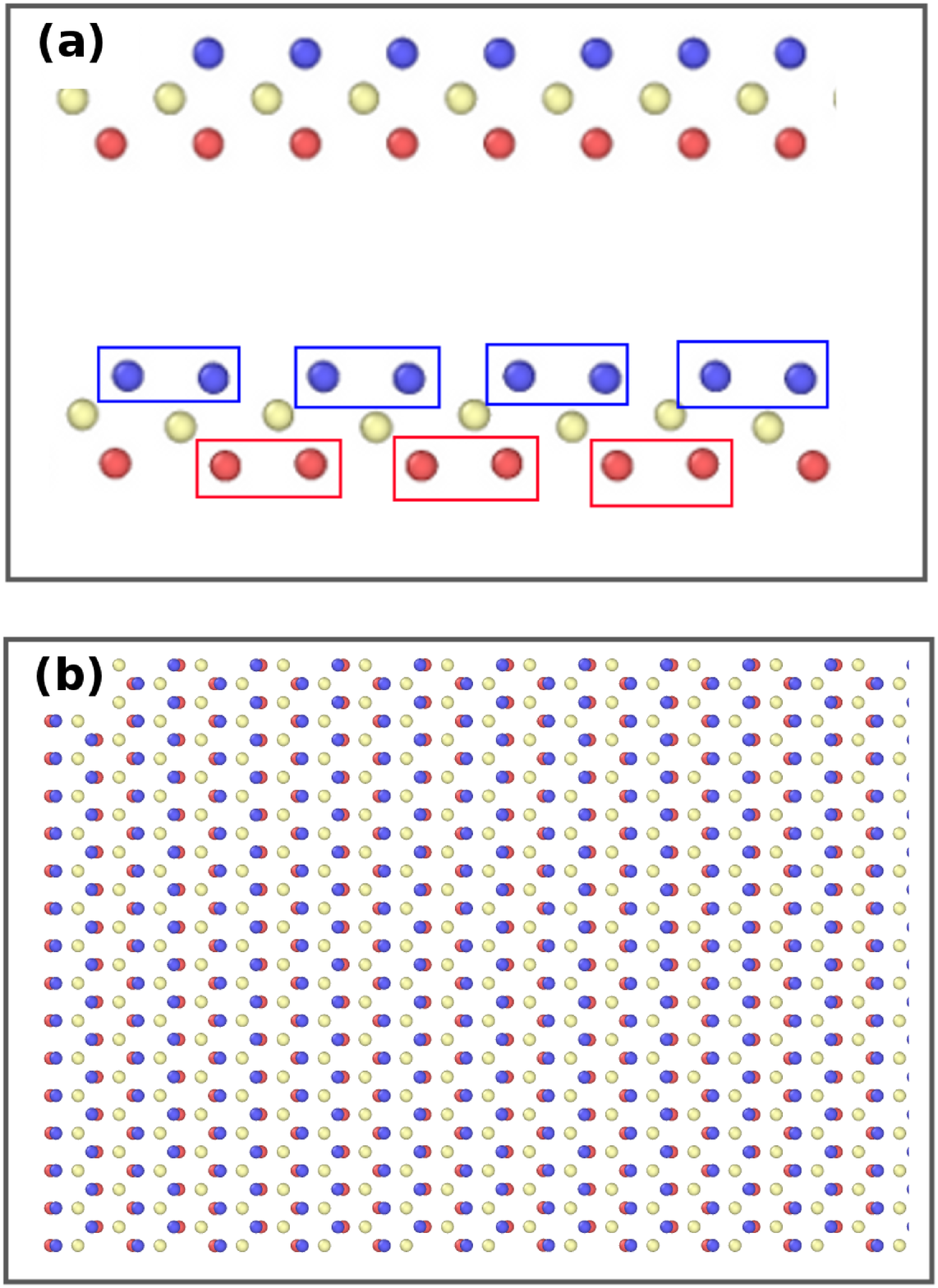}}
  \end{center}
  \caption{(Color online) Structure transition in the single-layer MoS$_{2}$ at tensile strain $\epsilon=0.193$ at 1.0~K. (a)  Side views of the single-layer MoS$_{2}$ before and after the structure transition. After transition, two neighboring S atoms construct a binary atomic pair (indicated by rectangular boxes). (b) Top view for the single-layer MoS$_{2}$ after structure transition.}
  \label{fig_cfg_mos2}
\end{figure}

An interesting mechanical response is that of a structural transition that occurs in both single-layer MoS$_{2}$ and the GMG heterostructure. Fig.~\ref{fig_stress_strain}~(a) shows a step-like jump around $\epsilon=0.19$ in the stress-strain curve for single-layer MoS$_{2}$, which is due to a structural transition involving a relative shift of the two outer S atomic layers, and which is related to the semiconductor-metallic phase transition in the single-layer MoS$_{2}$.\cite{GokiE2012acsn,LinYC2013,DangKQ2014sm} Fig.~\ref{fig_cfg_mos2} clearly demonstrates the relative shift of the outer S atomic layers during the structural transition. \jwj{Fig.~\ref{fig_cfg_mos2}~(a) shows the side views of the single-layer MoS$_{2}$ before and after the structure transition. After transition, two neighboring S atoms construct a binary atomic pair (indicated by rectangular boxes). The structure transition also induce some zigzag-like fluctuation for the position of the middle Mo atoms in the vertical direction. Before structure transition, S atoms in the side view are equally distributed in the horizontal direction. The space between two neighboring S atoms equals 3.1608~{\AA}. After structure transition, there are two different spaces, i.e 2.8019~{\AA} and 3.5852~{\AA}. The fluctuation of the position for the middle Mo atom layers is 0.3935~{\AA} in the vertical direction. Similar structure transitions are also observed for the MoS$_{2}$ in the GMG heterostructure. Fig.~\ref{fig_cfg_mos2}~(b) displays the top view of the material after structure transition.} This step-like jump becomes smoother at higher temperatures as shown in panels (b) and (c) because the outer two S atomic layers are already exhibiting larger and more frequent oscillations at higher temperature before the structural transition occurs. As a result, the influence introduced by the structural transition is overtaken by the thermal vibration in the stress-strain curve.

\begin{figure}[htpb]
  \begin{center}
    \scalebox{1.0}[1.0]{\includegraphics[width=8cm]{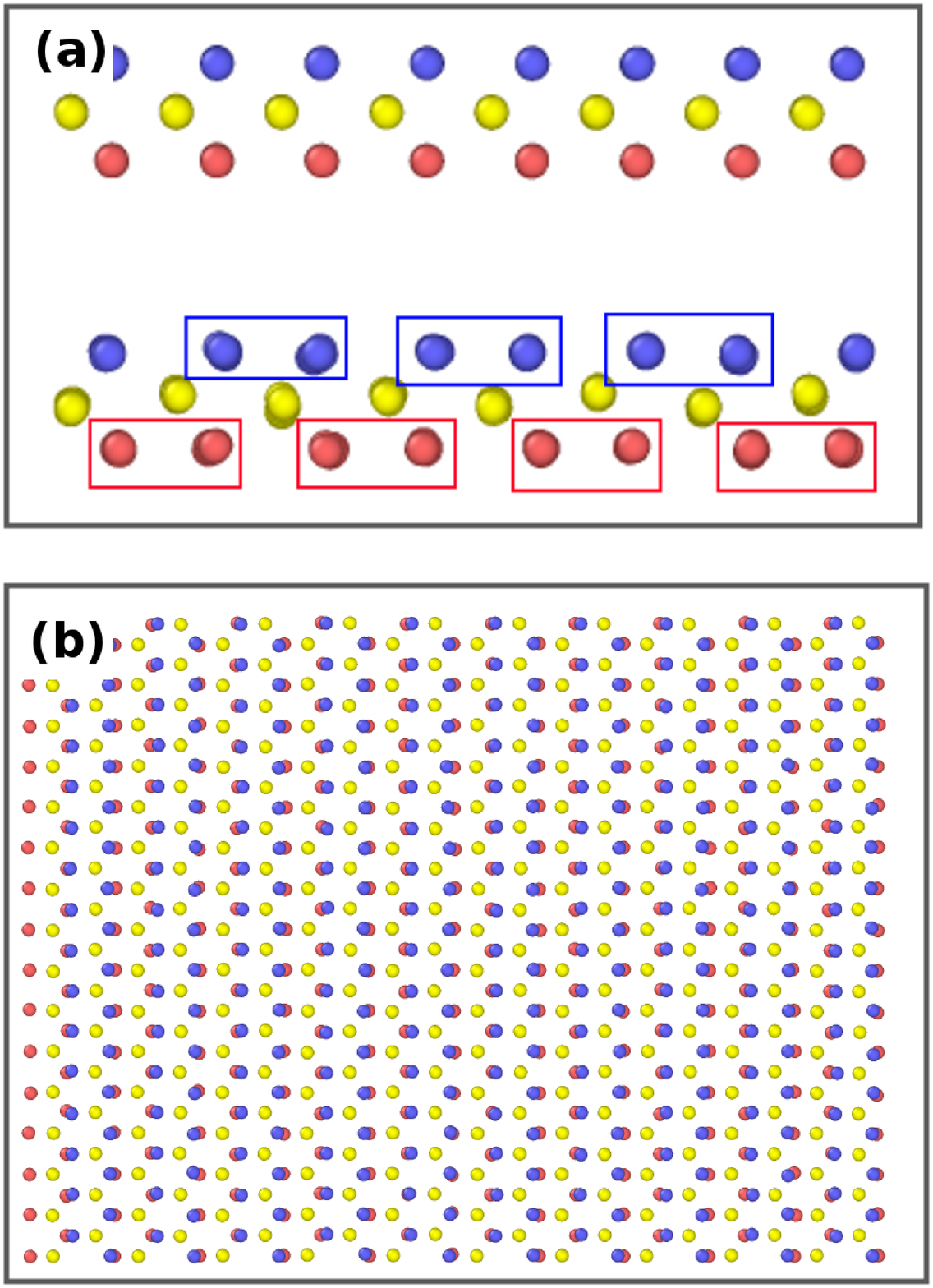}}
  \end{center}
  \caption{(Color online) Structure transition for the single-layer MoS$_{2}$ in the GMG heterostructure at tensile strain $\epsilon=0.193$ at 1.0~K. (a)  Side views of the MoS$_{2}$ before and after the structure transition. After transition, two neighboring S atoms construct a binary atomic pair (indicated by rectangular boxes). (b) Top view for MoS$_{2}$ in the GMG heterostructure after structure transition.}
  \label{fig_cfg_gmg}
\end{figure}

Fig.~\ref{fig_cfg_gmg} shows that the same structural transition happens in the MoS$_{2}$ layer sandwiched between two graphene layers. However, it happens that these two graphene layers yield at almost the same mechanical strain, so the step-like jump is concealed in the stress-strain curve for the GMG heterostructure. Instead, the stress in the heterostructure increases rapidly with increasing strain around $\epsilon=0.19$, disclosing the yielding phenomenon of the outer two graphene layers.

\begin{figure}[htpb]
  \begin{center}
    \scalebox{1.0}[1.0]{\includegraphics[width=8cm]{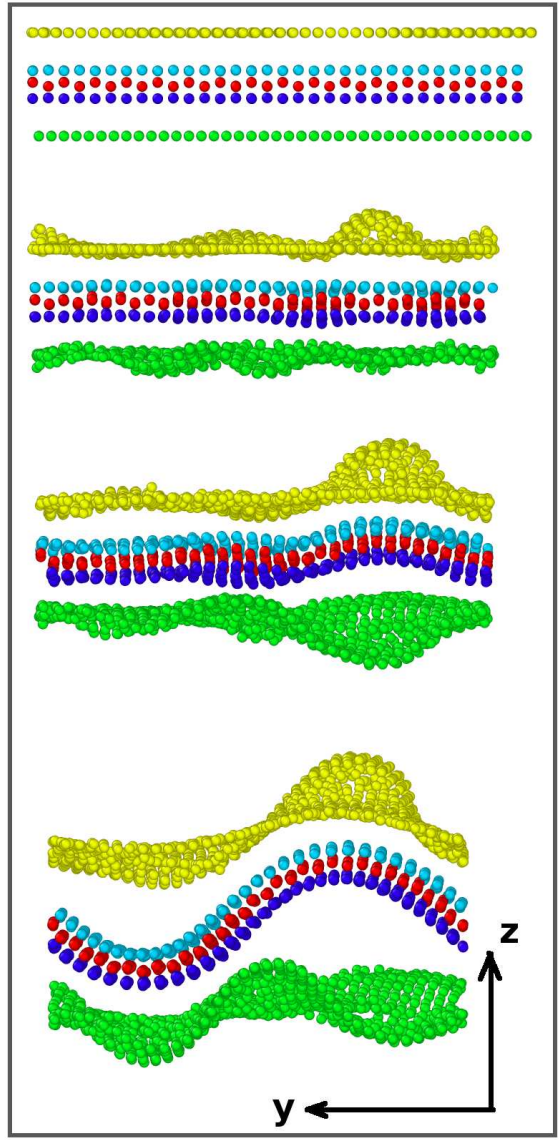}}
  \end{center}
  \caption{(Color online) Buckling of the GMG heterostructure at the ultimate strain $\epsilon=0.26$ at 1.0~K. GMG is stretched in the x direction, resulting in the compression in the y direction. From top to bottom, the tension in the x direction increases as 0.2613, 0.2614, 0.2616, and 0.2620.}
  \label{fig_cfg_gmg2}
\end{figure}

We now compare the difference in the ultimate strain of single-layer MoS$_{2}$ and the GMG heterostructure. Fig.~\ref{fig_stress_strain}~(a) shows that the ultimate strain in single-layer MoS$_{2}$ is around 0.4. \jwj{However, the ultimate strain in GMG heterostructure is much smaller than that of the single-layer MoS$_{2}$.} Fig.~\ref{fig_cfg_gmg2} displays the configuration for the GMG heterostructure during mechanical tension with strain close to the GMG ultimate strain $\epsilon=0.26$. These snapshots illustrate that the GMG heterostructure is compressed in the y direction when it is stretched in the x direction under external mechanical tension, \jwj{which is due to positive Poisson's ratio in both graphene and MoS$_{2}$.} \jwj{This compression in the y direction leads to the buckling of the two outer graphene layers. The second image shows that the inner MoS$_{2}$ layer is not buckled at this initial stage due to the fact that the bending modulus of single-layer MoS$_{2}$ is larger than that of graphene by a factor of seven.\cite{JiangJW2013bend}  After further tension, the inner MoS$_{2}$ layer also starts to buckle as influenced by the severe rippling exhibited by the sandwiching graphene layers.}

\begin{figure}[htpb]
  \begin{center}
    \scalebox{1.0}[1.0]{\includegraphics[width=8cm]{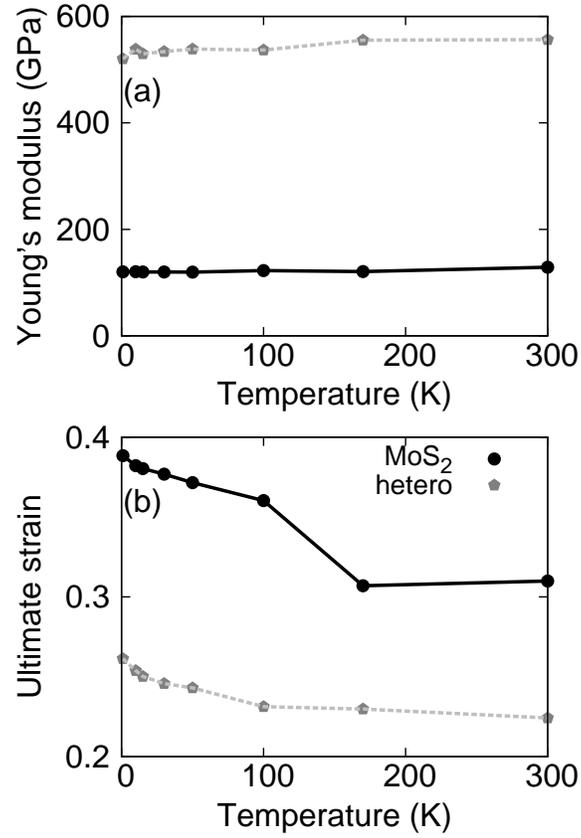}}
  \end{center}
  \caption{(Color online) Young's modulus and the ultimate strain in the MoS$_{2}$ and the GMG heterostructure. (a) Temperature dependence for the Young's modulus. (b) Temperature dependence for the ultimate strain in two systems.}
  \label{fig_temperature}
\end{figure}

Fig.~\ref{fig_temperature} compares the Young's modulus in single-layer MoS$_{2}$ and the GMG heterostructure at different temperatures. Fig.~\ref{fig_temperature}~(a) shows that the Young's modulus in the heterostructure is over 500.0~GPa, which is much higher than the pure MoS$_{2}$ layer. Fig.~\ref{fig_temperature}~(b) shows that the ultimate strain is reduced after the MoS$_{2}$ is sandwiched by two graphene layers. In both systems, the ultimate strain decreases with increasing temperature due to stronger thermal vibrations at higher temperature.

\section{conclusion}
To summarize, we have performed classic molecular dynamics simulations to comparatively investigate the mechanical properties of single-layer MoS$_{2}$ and the GMG heterostructure. We find that that the lattice mismatch between MoS$_{2}$ and graphene will result in an spontaneous intrinsic strain energy in the heterostructure. Our study shows that the GMG heterostructure exhibits a Young's modulus that is about three times that of single layer MoS$_{2}$, while correspondingly exhibiting a yield strain that is about 30-40\% smaller than that of single layer MoS$_{2}$.

\textbf{Acknowledgements} The work is supported by the Recruitment Program of Global Youth Experts of China and the start-up funding from Shanghai University. HSP acknowledges the support of the Mechanical Engineering department at Boston University.


%

\end{document}